\documentclass[twocolumn]{aastex63}

\newcommand{\logg}{\mbox{$\log g$}}
\newcommand{\teff}{\mbox{$T_{\rm eff}$}}

\shorttitle{Sample article}
\shortauthors{de Mijolla}
\graphicspath{{./}{figures/}}
\usepackage{amsmath}
\usepackage{graphicx}
\usepackage{todonotes}
\usepackage{comment}
\usepackage{float}
\usepackage[ruled,vlined]{algorithm2e}

\begin{document}

\title{Measuring chemical likeness of stars with RSCA}

\correspondingauthor{Damien de Mijolla}
\email{damiendemijolla@gmail.com}

\author{Damien de Mijolla}
\affiliation{Department of Physics and Astronomy,\\
University College London,
\\ Gower Street, WC1E 6BT, UK}

\author{Melissa Ness}
\affiliation{Department of Astronomy,\\ 
Columbia University, \\
Pupin Physics Laboratories, New York, NY 10027, USA}
\affiliation{Center for Computational Astrophysics, Flatiron Institute, 162 Fifth Avenue, New York, NY, 10010, USA}



\begin{abstract}
Identification of chemically similar stars using elemental abundances is core to many pursuits within Galactic archaeology. However, measuring the chemical likeness of stars using abundances directly is limited by systematic imprints of imperfect synthetic spectra in abundance derivation. We present a novel data-driven model that is capable of identifying chemically similar stars from spectra alone. We call this Relevant Scaled Component Analysis (RSCA). RSCA finds a mapping from stellar spectra to a representation that optimizes recovery of known open clusters. By design, RSCA amplifies factors of chemical abundance variation and minimizes those of non-chemical parameters, such as instrument systematics. The resultant representation of stellar spectra can therefore be used for precise measurements of chemical similarity between stars. We validate RSCA using 185 cluster stars in 22 open clusters in the APOGEE survey. We quantify our performance in measuring chemical similarity using a reference set of 151,145 field stars. We find that our representation identifies known stellar siblings more effectively than stellar abundance measurements. Using RSCA,  1.8\% of pairs of field stars are as similar as birth siblings, compared to 2.3\% when using stellar abundance labels. We find that almost all of the information within spectra leveraged by RSCA fits into a two-dimensional basis, which we link to [Fe/H] and $\alpha$-element abundances. We conclude that chemical tagging of stars to their birth clusters remains prohibitive. However, using the spectra has noticeable gain, and our approach is poised to benefit from larger datasets and improved algorithm designs. 

\end{abstract}



\section{Introduction} \label{sec:intro}

The field of Galactic astronomy has entered a transformative era. Large-scale surveys, such as APOGEE, GAIA and GALAH, are providing millions of high-quality spectroscopic and astrometric measurements of stars across the Milky Way \citep{Rave2006, Apogee,Galah,Lamost,Gaia}. Future large-scale surveys, which will release even more high-quality data, are on the horizon \citep{4most,SDSS5,Weave}.

In this landscape of high-volume high-quality stellar astronomy, fully extracting the scientifically relevant information from stellar spectra remains a difficult problem. Classically, this has been done by comparing observations to synthetic spectra generated from theoretical models \citep[e.g.][]{aspcap}. However, the precision with which stellar labels can be derived under such an approach is ultimately limited by the faithfulness with which synthetic spectra reproduce observations. Because of computational constraints and gaps in knowledge, synthetic spectra do not perfectly match observations, something sometimes referred to as the ``synthetic gap" \citep{cyclestar}. Computational models used to generate synthetic spectra use incomplete stellar line lists and usually must make simplifying assumptions. This includes for example that stellar atmospheres are one-dimensional, in hydrostatic equilibrium, and in local thermodynamic equilibrium. In addition, even beyond these issues, observations are affected by further systematics such as telluric lines introduced by the earth's atmosphere \cite[e.g.][]{SDSSdataproducts} and telescope imperfections/aberrations.

Ultimately this synthetic gap limits our ability to extract information from stellar spectra. In \cite{CramerRaoTing,HowManyElements} it was shown that stellar spectra contain more chemical information than is captured in bulk metallicity and $\alpha$-enhancement alone. The precision of derived individual stellar abundances from large surveys, however, may be limited by an inability to fully extract information given approximate models, rather than by the signal to noise of observations.

This is problematic because a lot of interesting science requires measuring chemical similarity between stars with a precision beyond that currently delivered by large stellar surveys modelling pipelines. In particular, high precision chemical measurements are needed for strong chemical tagging \citep{TaggingFreeman}. This is an ambitious Galactic archaeology endeavour aiming to identify stellar siblings - stars born from the same molecular cloud - using chemical information derived from spectroscopy long after clusters gravitationally dissipate. In practice, whether such chemical tagging is theoretically possible at scale is still an open question, but may be answered with large-scale surveys like GALAH \citep{Galah, Buder2021}. For this form of chemical tagging to be successful, stellar siblings must share a near-identical chemical composition with sufficient variability in chemical compositions between clusters. Even if strong chemical tagging, reveals itself to be impossible at large scale, precise chemical similarity measurements would still be useful in reconstructing the broad nature of our galaxy's evolution \citep[e.g.][]{Coronado2020, Kamdar2020}.

These issues motivate the development of  methods capable of extracting information from stellar spectra. and overcoming the synthetic gap between observations and theoretical models. Several data-driven methods have been developed for this purpose. Methods such as those proposed \cite{Ness2015,Casey2016,Leung2018,Ting2019,cyclestar,Das_2019} allow for improving precision of stellar labels through leveraging data-driven interpolators between stellar spectra and labels, reducing the impact of noise and systematics on derived parameters. However, as such approaches still rely on synthetic spectra they do not fully alleviate issues with systematic errors from mismatching theoretical spectra. Recently, methods for finding chemically similar stars directly from stellar spectra without reliance on synthetic spectra have been developed \cite{Bovy2016,pca_dimensionality,Cheng2020TestingTC,de_Mijolla_2021}. This category of method works by removing the effect of non-chemical parameters on stellar spectra, thus isolating the chemical information within the spectra. Such approach are not without drawbacks. Although they remove the dependency on  synthetic models, they still require a comprehensive and precise determination of all non-chemical factors of variation. Additionally, they must make simplifying assumptions regarding the cross-dependencies between chemical and non-chemical factors of variations which may have an impact on accuracy.

In this paper, we present a new approach for identifying chemically similar stars from spectroscopic data which we name ``Relevant Scaled Component Analysis" (RSCA) because of its similarities with Relevant Component Analysis \citep{relevantComponentAnalysis}. Our approach is grounded in the machine learning subfield of metric learning. Instead of estimating individual chemical abundances, we project spectra directly into a lower dimensional subspace in which distances between spectra are made to encode a useful notion of chemical similarity between stars. Crucially, as our approach for transforming stellar spectra does not rely at any stage on synthetic spectra or quantities derived from these, its performance is not hindered by inaccuracies in stellar modelling.

A novelty of our work is that instead of using synthetic spectra to learn this notion of chemical similarity, we make use of spectra from know open clusters with open cluster membership information. Open clusters are groups of stars born together that remain gravitationally bound after birth and up to the present day. They are relatively rare, as most stellar clusters dissipate rapidly after birth \citep{openclusterreview}. However, they are extremely useful tools in modern Galactic astronomy. In particular, open clusters, which can be identified using astrometry, display near-identical chemical abundances although small scatter may exist at the 0.01 to 0.02 dex level and up to $<$0.05 dex level for some elements \cite[e.g.][]{Bovy2016,AtomicDiffusionLiu, Ness2018,Cheng2020TestingTC}. Open clusters have found many uses in modern astronomy. For example to obtain high-precision measurements of the radial abundance gradients in the Milky Way \citep{abundancegradientold,abundancegradientnew} or to benchmark and calibrate stellar survey abundance measurements \citep{aspcap}. Here, we use open clusters as a golden standard for learning a notion of chemical similarity. In our approach, we take the viewpoint that if open clusters are indeed chemically homogeneous, then a successful metric for encoding chemical similarity will be one in which open cluster stellar siblings are highly clustered.

Our algorithm, RSCA, has several  properties that make it suitable for the task at hand, of measuring chemical similarity between stars:
\begin{itemize}
    \item It is fully \textit{data-driven}. Chemical similarity is measured without any reliance or dependency on theoretical models. This offers a measure of chemical similarity that is independent from the systematics introduced in traditional stellar modelling \citep[e.g.][]{Jofre2017}, and offers a means of validating existing discoveries.
    
    \item It is \textit{computationally efficient}. As the method is linear, processing spectra from the full APOGEE stellar survey can be done in minutes. The most computationally intensive step of the approach is  a Principal Component Analysis decomposition.
    
    \item It is \textit{interpretable}. In its current formulation, measuring chemical similarity using our method amounts to evaluating Euclidean distances between stars projected on a hyperplane of the stellar spectra space.
    
    \item It is \textit{precise}. We find the method, using spectra, to be more effective at identifying stellar siblings from open clusters than is possible using stellar abundance measurements. We believe this to be in large part because our method bypasses the synthetic gap introduced by spectral modelling. Furthermore, our experiments suggest that the performance could be further improved, for example, with a larger dataset of open cluster stars or by taking into account the error on the flux which we do not currently do.
\end{itemize}

The paper is organized as follows. In Section \ref{sec:metric}, we outline the conceptual ideas behind our approach for measuring chemical similarity. We then briefly introduce Principal Component Analysis in Section \ref{sec:PCA}, which is a core component of our algorithm. In Section \ref{sec:algo}, we dive deeper, and present our algorithm, RSCA. This is implemented using open clusters observed by the APOGEE survey in Sections \ref{sec:experiments}, and evaluated in light of the field distribution of stars. Its trade-offs and implications are discussed in Section \ref{sec:discussion}.

\section{Concepts and Assumptions}

\subsection{Chemical similarity as metric learning}\label{sec:metric}

The characteristics within a stellar spectrum are caused by the interplay of many factors of variation. These include chemical and physical parameters of the star and the instrumental systematics associated with the telescope, as well as interstellar dust along the line of sight. Measuring chemical similarities requires disentangling the imprint left on the spectra by chemical factors of variation from that left by the other non-chemical factors of variation. Our goal is to identify chemically similar stars from their spectra, for stars that span a range of physical stellar parameters (i.e. effective temperatures and surface gravities). We approach this task from a data-driven perspective, and build an algorithm for identifying stars that are as chemically similar as birth siblings, using open cluster spectra.

For our method, we assume that open clusters are close to chemically homogeneous because of their common birth origin \citep{Ness2018} but are not special in any other way (at least in terms of their spectra). That is to say, we assume that the only information within spectra useful for recognizing open clusters are the chemical features of the spectra, and so that a model which identifies open clusters from the spectra will need to do so by extracting the chemical information within spectra

We frame the task of building such a model recognizing open clusters as a metric learning task. That is to say, we build a data-driven model converting stellar spectra into a representation in which Euclidean distances convey the uncalibrated probability of stars originating from a shared open cluster. To accomplish this, the training objective of our data-driven algorithm can be understood as transforming stellar spectra into a representation in which the distance between intra-cluster stars is minimized and the distance between inter-cluster stars is maximized.

Distances in the representation resulting from such an optimizing procedure will organically quantify the chemical similarity of stars. Non-chemical factors of variation, such as stellar temperatures and instrumental systematics, will not contribute to the representation as their presence would make distances between stellar siblings larger. Instead, such a representation will only contain those factors of variation of spectra that are discriminative of open clusters, ie the chemical factors of variation.  Crucially, chemical factors of variation will contribute to distances in the representation in proportion to how precisely they can be estimated from stellar spectra. Stronger chemical features will be more strongly amplified than weaker chemical features.

The utility of this data-driven approach is that it is independent of imperfect model atmosphere approximations and other issues affecting synthetic spectra. This provides a high fidelity technique to turn to specific applications within Galactic archaeology, such as chemical tagging of stars that are most chemically similar \citep{TaggingFreeman}.

The assumption underpinning our work is that this chemical information will be the only information within stellar spectra useful for distinguishing open clusters and so will be the only information captured by our model. If this assumption is true, then the representation induced by the model will measure a form of chemical similarity between stellar spectra.

However, since open clusters, in addition to sharing a common age and near-identical birth abundances, are also gravitationally bound, they can be identified from their spatial proximity if such information is available in the spectra. As such spatial information does not robustly transfer towards identifying dissolved clusters, it mustn't  be captured by our model. In this work, we apply our algorithm to pseudo-continuum normalized spectra with diffuse interstellar bands masked, which we assume not to contain any information about spatial location so as for our representation after training to only contain chemical information. Assuming pseudo-continuum normalized spectra do not encode any spatial information is plausible, since after continuum normalization, the spectrum should not contain significant information about stellar distance. With the impact of reddening removed and diffuse interstellar bands masked, a spectrum should also not contain any information about the stellar extinction and interstellar medium along the line of sight of the star. We examine the validity of these assumptions in latter sections. Ultimately, it is worth emphasizing that our method only exploits features proportionally to their discriminativeness at recognizing open clusters. Therefore, we can expect our model to not heavily rely on non-robust features (provided that these are significantly less informative than robust features).  This also relies on our open cluster training data being representative of the parameter space that should be marginalized out; i.e. our model does not learn to associate inter cluster stars via \logg\ and \teff, which could happen if the evolutionary state of observed cluster stars was similar within clusters and different between clusters.

\begin{figure*}
\includegraphics[width=\linewidth ,page=7,trim={2cm 0cm 1cm 0cm},clip]{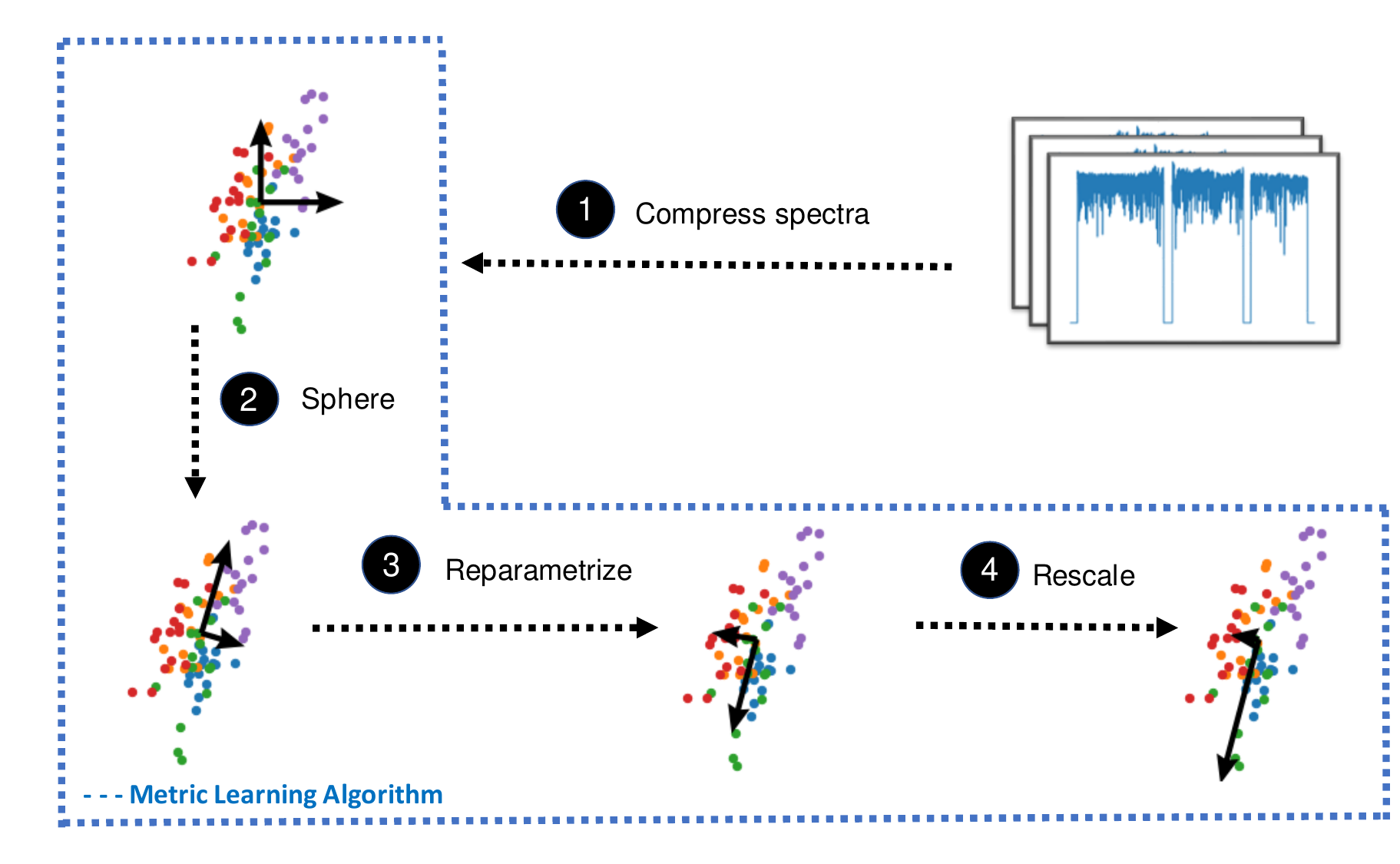}
\caption{Schematic depiction of RSCA.  The algorithm proceeds by first encoding stellar spectra into a lower dimensional representation made two-dimensional for illustrative purposes. In this representation, stellar siblings - which are represented by same-coloured dots - are not initially identifiable by their Euclidean distance in the basis (represented by black arrows). The objective of the metric-learning algorithm (dashed blue) is to find a new basis in which distances are informative about which stars are stellar siblings. This objective is realized through three linear steps: a sphering transformation on the dataset, a reparametrization to a suitable basis, and a scaling of the basis vectors.}
\label{fig:diagram_algorithm}
\end{figure*}

\subsection{Principal Component Analysis}\label{sec:PCA}
Our metric-learning algorithm, RSCA, first uses Principle Component Analysis (PCA) to transform the data into a (lower-dimensional) basis that represents the primary variability of the ensemble of spectra we work with.

The principal components of a dataset $X$, of shape $N_{D} \times N_{F}$ containing $N_{D}$ data points and $N_{F}$ features, are an ordered orthogonal basis of the feature space with special properties. In the principal component basis, basis vectors are ordered by the amount of variance they capture. They have the property that for any k, the hyperplane spanned by the first k-axes of the basis is the k-dimensional hyperplane, which maximally captures the data variance. In PCA, the number of principal components used, k, is a hyperparameter controlling the trade-off between the amount of information preserved in the dataset $X$ after compression and the degree of compression.

The principal component basis corresponds to the unit-norm eigenvectors of the covariance matrix of $X$ ordered by eigenvalue magnitude. This can be obtained through diagonalization of the covariance matrix. The principal component basis can also be formulated as the maximum likelihood solution of a probabilistic latent model which is known as Probabilistic Principal Component Analysis (PPCA) (see \cite{bishop}). This probabilistic formulation is useful in that it enables one to obtain the principal components for a dataset containing missing values by marginalizing over these.

As we will make use of further in this paper, the principal components also allow for generating a sphering transformation. This is a linear transformation of the dataset to a new representation, in which the covariance matrix of the dataset $X$ is the identity matrix. Sphering using PCA is carried out by performing a change-of-basis to a modified principal component basis in which the principal components are divided by the square root of their associated eigenvalues.

\section{Relevant Scaled Component Analysis Algorithm}\label{sec:algo}

The input to RSCA is individual stellar spectra, some of which belong to open clusters, and some of which are a reference field sample. The output of RSCA is, for each spectra, a $N_{k}$ vector, in which dimensions are scaled such that distances between  $N_{k}$ vectors of pairs of stars encode chemical similarity between those stars.
We step through this in detail below.

\subsection{Overview}

Let us define $X_{\rm clust}$ as the matrix representation of a dataset containing the spectra of known open cluster stars. Analogously, let us define $X_{pop}$ as the matrix representation of a larger dataset of stellar spectra in the field (with unknown cluster membership). These matrices are respectively of shapes $N_{\rm d_{clust}} \times N_{\rm b}$ and $N_{\rm d_{pop}} \times N_{\rm b}$, where $N_{\rm d_{clust}}$ is the number of open cluster stars, $N_{\rm d_{pop}}$ the number of of stars in the large dataset and $N_b$ the number of spectral bins. For our purposes, we assume access to only a limited number of open cluster stars such that $N_{\rm d_{clust}} < < N_{\rm d_{pop}}$. We also assume that the spectra in these matrices are pseudo-continuum normalized spectra, with diffuse interstellar bands masked in a process following that described in Appendix \ref{appendix:masking}. Pseudo-continuum spectra are normalized rest-frame spectra in which the effects of interstellar reddening and atmospheric absorption are removed, in a process described in \cite{Apogee}.

RSCA takes as inputs $X_{clust}$ and $X_{pop}$. Through a series of linear-transformation to these matrices, RSCA maps these matrices into new matrices of shape $N_{\rm d_{clust}} \times N_{\rm K}$ and $N_{\rm d_{pop}} \times N_{\rm K}$ whose entries are the stellar spectra transformed to a metric-learning representation of dimensionality $K$. Euclidean distances in this new metric-learning representation can then be used to measure chemical similarity between spectra. As all the steps of RSCA are linear transformations, the mapping converting from spectra to the metric learning representation can be parameterized by an $N_{\rm K} \times N_{\rm bins}$ matrix and used to convert unseen spectra (or for visualization purposes).

We provide in Figure \ref{fig:diagram_algorithm} a graphical depiction of the linear transformations involved in the RSCA algorithm. RSCA works by first projecting the spectra onto a set of basis vectors with PCA (Step 1). For visualization purpose, this basis is made two-dimensional although it would normally be higher dimensional. After this PCA compression (Step 1), stellar siblings are represented as same-coloured dots whose xy coordinates correspond to coordinates in the PCA basis. Once in the PCA basis a series of linear transformation are applied to the data. For improved clarity, we keep the data fixed throughout our algorithm visualization and represent linear transformations as change-of-basis (black arrows). Step 2 and 3 of the algorithm, find a new basis which more aptly captures spectral variability amongst stellar siblings and Step 4 of the algorithm rescales basis vectors of this basis based on a comparison between their spectral flux variance amongst stellar siblings and amongst field stars. The outcome of the RSCA algorithm is a new representation of the spectra in which dimensions that are unhelpful for discriminating stellar siblings are minimised in amplitude  (through a stretching-out of basis vectors). Conversely, dimensions that are helpful in recognizing stars within the same open clusters are made larger (through a squeezing of basis vectors). The $K$ vector of principle components for each star can be collapsed into a measure of chemical similarity through a Euclidean distance measure between stars of their scaled representation, output from RSCA, where $d$ = $\sqrt{(n_{K}- n'_{K})^2}$ for any pair of stars $n,n'$.

We now walk step-by-step through the successive linear-transformations involved in the RSCA algorithm. For following along, the pseudo-code for RSCA is provided in Appendix \ref{appendix:pseudocode} and the full source code of our project which contains a Python implementation is made available online \footnote{https://github.com/drd13/RSCA}.

\subsection{Step 1: Compress the spectra with PCA to reduce the risk of overfitting}

In the first step of our approach, denoted as \textbf``1) Compress Spectra", in Figure \ref{fig:diagram_algorithm}, we apply PCA to $X_{pop}$ to convert the population of stellar spectra into a lower dimensional representation. This dimensionality reduction step serves to make the algorithm more data-efficient which is crucial given the risk of overfitting from the small number of open clusters within our dataset.

As some stellar bins are flagged as untrustworthy, we use Probabilistic PCA (PPCA), a variant of the PCA algorithm which can accommodate missing values. After finding principal components of $X_{pop}$, we compress the data by discarding all but the $K$ largest principal components, where $K$ is a hyperparameter requiring tuning. Then,  datasets $X_{pop}$ and $X_{clust}$ are each transformed using the $K$ basis vectors, which we call $Z_{pop}$ and $Z_{clust}$; the representation of the spectra in the PCA basis of $X_{pop}$. These haves shapes of $N_{\rm d_{pop}} \times N_{\rm K}$ and $N_{\rm d_{clust}} \times N_{\rm K}$.

\subsection{Metric Learning: Sphering, Reparameterization and Rescaling}

Step (1) in our procedure of our PCA compression is a pre-processing step. The steps that follow fall into the realm of a general-purpose metric-learning algorithm. These rely on assumptions about the PCA-compressed spectra being satisfied. Performance should be robust to within small departures from these assumptions, but will still ultimately be tied to how well these assumptions are respected. We lay out our assumptions for steps (2)-(4) below.

\subsubsection{Assumptions}

First, we assume that the data (i.e. spectra) in $Z_{pop}$ are well approximated as being drawn from a multivariate Gaussian distribution. That is to say that if we define $\mu_{pop}$ and $\Sigma_{pop}$ as the mean and covariance of $Z_{pop}$, then the stars within $Z_{pop}$ can be assumed as being samples drawn from $z_{pop} \sim N(\mu_{pop},\Sigma_{pop})$.

Next, we make the assumption that individual clusters are themselves approximately Gaussian in the PCA-compressed space. That is to say we posit that the members of open clusters are well approximated as being samples drawn from a distribution $z_{clust} \sim N(\mu_{clust},\Sigma_{clust})$. Crucially, we make the assumption that all open clusters share the same covariance matrix $\Sigma_{clust}$ and only differ in their mean $\mu_{clust}$. This is perhaps our strongest and most important assumption. That is, that the stars within different clusters are distributed following a shared covariance matrix (i.e. clusters have the same shape irrespective of their location in the representation). It is this assumption which allows a linear transformation i.e. a transformation that acts the same across the whole representation, to be an effective approach for  measuring chemical similarity. This assumption of what is effectively cluster translation invariance, can be interpreted as assuming that the scatter amongst stellar spectra in physical and chemical parameters should be the same for all clusters irrespective of the clusters parameters which is a sensible assumption. Connecting these assumptions back to Figure \ref{fig:interpretation}, each step 2-4 requires that stars in any population, within clusters and within the field, follow a multivariate Gaussian with an invariant covariance matrix for each individual cluster. 

\subsubsection{Step 2: Sphering to transform the population covariance matrix into the identity matrix}

Together, the sphering and reparametrization steps of RSCA  serve to transform $\Sigma_{pop}$ and $\Sigma_{clust}$ to a vector representation of stellar spectra in which $\Sigma_{pop}$ and $\Sigma_{clust}$, the covariance matrices of the field stars and clusters, respectively, are diagonal matrices. As there are then no off-diagonal terms in the covariance matrices, this ensures that the variances along basis vectors fully capture the covariance information amongst stellar siblings and amongst field stars.

 In Step \textbf{``2) Sphere"} of our algorithm, in Figure \ref{fig:diagram_algorithm}, we  linearly transform the vector representation of spectra such that the dataset $Z_{pop}$ has a covariance matrix of identity after transformation. This linear transformation takes the form of a sphering transformation applied to $Z_{pop}$, where in our experiments we use the PCA-sphering scheme. This has the utility of fully capturing the variability amongst stellar spectra and amongst stellar siblings.

\subsection{Step 3: Reparameterization to diagonalize the cluster covariance matrix}

In steps 1 and 2 of RSCA we do operations on the full field population and full cluster population, respectively. In steps 3 and 4, we transform and scale to recognise stellar sibling likeness compared to the field, so consider stellar variability of the stars within individual clusters individually.

In the next step of our algorithm after the sphering transform, \textbf{``3) Reparametrize"} in Figure \ref{fig:diagram_algorithm} represents a change of basis. 
In this step, the covariance matrix $\Sigma_{clust}$, is diagonalized. Since we do not have direct access to $\Sigma_{clust}$, as an intermediary step in doing so, a dataset approximately distributed according according to $N(0,\Sigma_{clust})$ is created from the open cluster dataset. This is done by subtracting each star's vector representation from the mean representation of all stars belonging to the same cluster $\widehat{\mu}_{clust}$, such that each cluster becomes zero-centered. It is worth noting that as $\widehat{\mu}_{clust}$ is estimated from a limited number of samples, it will not exactly match with the true $\mu_{clust}$ and so the resultant population will only approximately be distributed according to $N(0,\Sigma_{clust})$. 

As PCA basis vectors correspond to (unit-norm) eigenvectors of covariance matrices, the PCA basis obtained by applying PCA to the zero-centered open cluster dataset parametrizes a transformation to a representation in which $\Sigma_{clust}$ is (approximately) diagonal. A change-of-basis to this PCA basis thus parametrizes the desired diagonalization of the cluster covariance matrix. Because the basis vectors of the PCA basis have by construction unit-norm, the covariance matrix $\Sigma_{pop}$ will still be the identity matrix after this change-of-basis and hence both $\Sigma_{clust}$ and $\Sigma_{pop}$ will be diagonal matrices as desired.

\subsection{Step 4: Scaling to maximise discriminative power in identifying chemically similar stars}\label{sec:scaling}

In the final step of the metric learning algorithm, denoted by \textbf{``4) Rescale"} in Figure \ref{fig:diagram_algorithm}, basis vectors are scaled proportionally to their dimension's usefulness at recognizing open clusters. This is done by applying separately to each dimension of the representation a scaling factor. Here, the independent scaling of dimensions is justified by $\Sigma_{clust}$ and $\Sigma_{pop}$ being diagonal covariance matrices.

We use the separate individual clusters within $X_{clust}$ and field populations $X_{pop}$, to measure variance in each dimension to determine our scaling factor. We design our scaling factor such that, after transformation, distances between pairs of random stars quantify the ratio between the probability that pairs of stars originate from the same (open) cluster and the probability that they do not. To put it another way, we seek to scale dimensions such that pairs of stars that are more likely to originate from the same cluster as compared to originating from different clusters have a smaller separation (i.e. Euclidean distance) in the representation than less likely pairs.

Under our set of assumptions, along a dimension $i$ amongst the K dimensions, random stellar siblings (intra-cluster stars) are distributed as $z_{clust} \sim N(\mu_{clust_i},\sigma_{clust_i})$ where $\mu_{clust_i}$ and  $\sigma_{clust_i}$ are the mean and standard deviation along dimension $i$ (at this stage in the algorithm).  Accordingly, using the standard formula for the sum of normally distributed variables, the one-dimensional distance along $i$ between pairs of random stellar siblings $z_{clust_{1i}}$ and $z_{clust_{2i}}$ follows a half-normal distribution $d_{clust_i} \sim |z_{clust_{1i}}-z_{clust_{2i}}|=|N(0,2 \sigma_{\rm clust_i})|$. Likewise, the distance $d_{pop_{i}}$ between pairs of random field stars follows a similar half-normal distribution $d_{pop_i} \sim  |N(0,2 \sigma_{\rm pop_i})|$ where $\sigma_{\rm pop_i}$ is the standard deviation amongst field stars along dimension $i$.

For a pair of stars observed a distance $d_{i}$ away from each other along a dimension $i$, the ratio between the probability of the pair originating from the same cluster (intra-cluster) and the probability of the pair not originating from the same cluster (inter-cluster) is:

\begin{equation}
r_{i}(d_i)=\frac
{
p(d_{clust_i}=d_{i})
}
{
p(d_{pop_i}=d_{i})
}=
|
\frac
{
N(0,2 \sigma_{\rm clust_i})
}
{
N(0,2 \sigma_{\rm pop_i})
}|
\end{equation}

which evaluates to (as distances $d_{i}$ are by design greater than 0)

\begin{equation}
r_{i}(d_i) = A_i e^{\frac{-d_{i}^{2}}{2\sigma_{r_{i}}^{2}}}
\end{equation}

where 

\begin{equation}
    A_i=\frac{\sigma_{pop_i}}{\sigma_{clust_i}}
\end{equation} and 
\begin{equation}
    \sigma_{r_{i}}= \frac{\sigma_{\rm clust_{i}}\sigma_{\rm pop_{i}}}{\sqrt{\sigma_{\rm pop_{i}}^2-\sigma_{\rm clust_{i}}^2}}
\end{equation}

As dimensions are assumed to be independent, the probability ratio accounting for all dimensions is the product of the probability ratio of the separate dimensions:

\begin{equation}
r = 
\prod_{i=0}^{K}
\frac
{
N(0,2 \sigma_{\rm clust_i})
}
{
N(0,2 \sigma_{\rm pop_i})
} 
= 
Ce^{
-\frac{1}{2}
\sum_{i=0}^{D} 
(\frac{d_i}{\sigma_{r_i}})^2
}
\end{equation}

where $C =\prod_{i=0}^{K}A_{i}$.

From this expression, it can be seen that multiplying dimensions by a scaling factor of  $\frac{1}{\sigma_{r_i}}$ leads to a representation in which the probability ratio $r$ is a function of Euclidean distance and where pairs of stars with smaller Euclidean distance separation have a higher probability of originating from the same open cluster as compared to their probability of originating from different cluster than pairs with larger Euclidean separation. 

It is clear that dividing dimensions by a scaling factor of $\frac{1}{\sigma_{r_i}}$ induces a representation where distances $d$ which are measured between the scaled reconstructed representation. directly encode the probability of stars originating from the same cluster, as desired for our metric-learning approach. However, using this expression as a scaling factor requires evaluating the $\sigma_{clust_i}$'s and $\sigma_{pop_i}$'s along all dimensions. Because the representation has been sphered, the population's standard deviation is unity along all directions ($\sigma_{pop_i}=1$). We estimate the intra-cluster standard deviations using a pooled variance estimator:

\begin{equation}
    \sigma_{\rm clust_i}^{2}
    =\frac{
    \sum_{j=1}^{k}\left(n_{j}-1\right) \sigma_{ji}^{2}
    }{
    \sum_{j=1}^{k}\left(n_{j}-1\right)
    }
\end{equation}

where $\sigma_{ji}^{2}$ refers to the sample variance along dimension i for the sample of stars belonging to an open cluster j containing $n_{j}$ stars in $X_{clust}$. To make the algorithm more robust to the presence of any outliers in the dataset, such as misclassified stellar siblings, we use the median absolute deviation (MAD) as an estimator for the sample standard deviation $\sigma_{ji}$. That is 
\begin{equation}
    \mathrm{MAD}=\operatorname{median}\left(\left|X_{i}-\tilde{X}\right|\right)
\end{equation}
where $X_{i}$ and $\tilde{X}$ are respectively the data values and median along a dimension.

To better understand the effect of our scaling factor on the representation it is applied to, it is instructive to look into the impact it has on the distances between stars along dimensions of a representation. When stars belonging to the same cluster have, along a dimension, a similar standard deviation to the full population of stars (i.e. $\sigma_{clust} \approx \sigma_{pop}$), the dimension carries no information for recognizing cluster member stars and the scaling factor accordingly fully suppresses it $\sigma_{r} \rightarrow \infty$. On the other hand, for dimensions where the population's standard deviation $\sigma_{pop}$ is significantly larger than the cluster standard deviation $\sigma_{clust}$, the population's standard deviation is no longer relevant and $\sigma_{r} \approx \sigma_{clust}$. That is to say the scaling devolves into measuring distances relative to the number of standard deviations away from the cluster standard deviation.

\section{Experiments on APOGEE Data}\label{sec:experiments}

We validate our approach for encoding chemical similarity by testing its performance on real data obtained by the APOGEE survey data release 16 \citep{ApogeeDR16}. The APOGEE survey \citep{Apogee} is an infrared, high resolution, high signal-to-noise spectroscopic survey. The APOGEE survey uses a 300-fiber spectrograph \citep{APOGEEspectrographs} installed at the Sloan Digital Sky Survey telescope located at Apache Point Observatory \citep{ApachePointObservatory}. 

\subsection{Dataset Preparation}

For our experiments we use spectra from the public APOGEE data release DR16 \citep{ApogeeDR16} to create $X_{clust}$ and $X_{pop}$, our datasets of field and open cluster stars. Our field dataset $X_{pop}$ contains spectra for 151,145 red-giant like stars matching a set of quality cuts on the 16th APOGEE data release described below. Our open cluster dataset $X_{clust}$ contains spectra for 185 stars distributed across 22 open clusters,  obtained after further quality cuts using the OCCAM value-added catalogue \cite{Donor_2020}, a catalogue containing information about candidate open cluster observed by APOGEE. We also create baseline datasets $Y$ and $Y_{clust}$ containing stellar abundances for the stars in $X$ and $X_{clust}$. We include abundances for 21 species in $Y$ and $Y_{clust}$: C, CI, N, O, Na, Mg, Al, Si, S, K, Ca, Ti, TiII, V, Cr, Mn, Fe, Co, Ni, Cu, Ce. These abundances are derived from the X\_H entry in the allStar FITS file.

To create the dataset of field stars $X_{pop}$, we make the following dataset cuts. With the intention of only preserving red-giant stars, we discard all but those stars for which $4000<\teff<5000$~K and $1.5<\logg<3.0$~dex,  where we use the $\teff$ and $\logg$ derived by the ASPCAP pipeline. In addition, we further exclude any stars for which some stellar abundances of interest were not successfully estimated by the ASPCAP pipeline by removing any star containing abundances set to -9999.99 for any of our 21 species of interest. We also exclude all spectra for which the STAR\_BAD flag is set in ASPCAPFLAG. The pseudo-continuum spectra of those remaining stars, as found in the AspcapStar FITS file, were used to create the matrix $X_{pop}$ in which each column contains the spectra of one star.

To create the dataset of open cluster member stars $X_{clust}$, we cross-match our filtered dataset with the OCCAM value-added catalogue \cite{Donor_2020} so as to identify all candidate open clusters observed by APOGEE. We only keep those spectra of stars with open cluster membership probability CG\_PROB$>$0.8 \citep{gaudinclusters}. After this cross-match, we further filtered the dataset by removing those clusters containing only a single member stars as these are not useful for us. Additionally, we further discard one star with Apogee ID ``2M19203303+3755558" found to have a highly anomalous metallicity. After this procedure, 185 OCCAM stars remain, distributed across 22 clusters. We do not cut any stars based on their signal-to-noise ratio. The stars in $X_{pop}$ have a median signal-to-noise ratio of 157.2 and interquantile range of 102.0-272.4 while those in $X_{clust}$ have a median signal-to-noise ratio of 191.4 and interquantile range of 117.7-322.6. 

Because of cosmic rays, bad telluric line removal or instrumental issues, the measurements for some bins of stellar spectra are untrustworthy. We censor such bad bins to prevent them from impacting our low-dimensional representation. Censored bins are treated as missing values in the PPCA compression. In this work, we have censored any spectral bin for which the error (as found in the AspcapStar FITS file error array) exceeds a threshold value of 0.05. Additionally, we censor for all stars in the dataset, those wavelength bins located near strong interstellar absorption features. More detail about the model-free procedure for censoring interstellar features can be found in Appendix \ref{appendix:masking}.

\subsection{Measuring Chemical Similarity}\label{sec:measuringchemicalsimilarity}

Evaluating how good a representation is at measuring the chemical similarity of stars requires a goodness of fit indicator for assessing the validity of its predictions. We use the "doppelganger rate" as our indicator. This is defined as the fraction of random pairs of stars appearing as similar or more similar than stellar siblings according to the representation, where similarity is measured in terms of distance $d$ in the studied representation. It is worth noting that this procedure for estimating doppelganger rates is related but different from the probabilistic approach presented in \cite{Ness2018}. 

We estimate doppelganger rates on a per-cluster basis by measuring distances between pairs of stars in the RSCA output representation. For each cluster in $X_{clust}$, the doppelganger is calculated as the fraction of pairs composed of one cluster member and one random star whose distance in the studied representation, $d_{intra-family}$ is $\leq$ than the median distance amongst all cluster pairs $d_{inter-family}$ . That is, $d_{intra-family}$ are pairs composed of two confirmed cluster members within $X_{clust}$ and $d_{inter-family}$ is a pair with one random field star selected from $X_{pop}$ and one cluster member from the studied cluster in $X_{clust}$. When calculating $d_{inter-family}$, we only consider pairs of stars with similar extinction and radial velocity, that is to say with $\Delta \rm AK\_TARG<0.05$ and $\Delta \rm VHELIO\_AVG<5$. By only comparing stars at similar extinction and similar velocity, we ensure that any model being investigated cannot reduce it's doppelganger rate through exploiting extinction or radial velocity information in the spectra.

So as to facilitate comparisons between different representations, we aggregate the per-cluster doppelganger rates into a ``global" doppelganger rate which gives an overall measurement of a representations effectiveness at identifying open clusters. The global doppelganger rate is obtained by averaging the per-cluster doppelganger rates through a weighted average in which clusters are weighted by their size in $X_{clust}$.

There is an added subtlety to assessing a representation through its global doppelganger rate. There are very few open cluster stars in the dataset. Therefore, RSCA as a data-driven procedure applied to open clusters, is susceptible to overfitting to the open cluster dataset. To prevent overfitting from affecting results, we carry out a form of cross-validation in which clusters are excluded from the dataset used for the derivation of their own doppelganger rate. In this scheme, calculating the global doppelganger rate of an RSCA representation requires repeated application of our algorithm, each time on a different subset with one cluster removed, as many times as there are open clusters.

We caution that our cross-validation approach has some implications on the derived doppelganger rates. Because, every cluster's doppelganger rate is evaluated on a slightly different data subset, the quoted distances and doppelganger rates are not comparable from cluster to cluster. 

\subsection{PCA Dimensionality}

\begin{figure}
\includegraphics[width=\columnwidth]{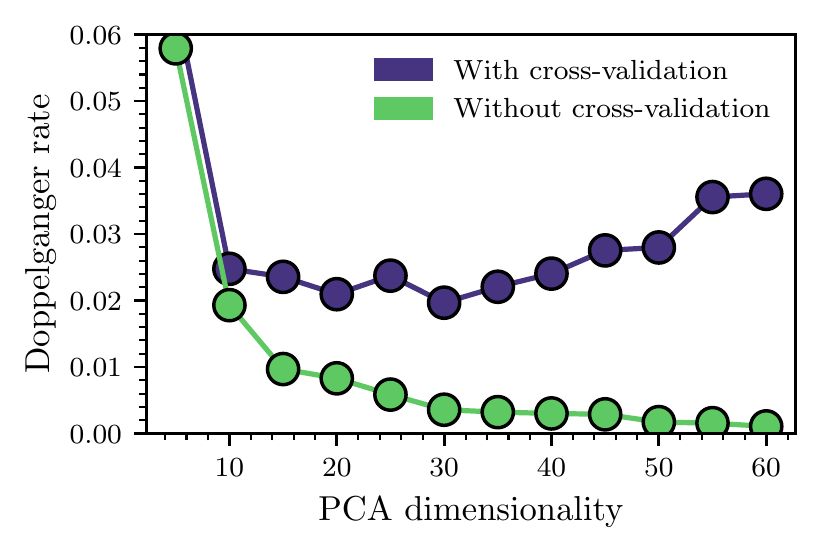}
\caption{Global doppelganger rates as a function of the number of PCA components used to encode spectra. Performance with cross-validation is shown in blue while performance without cross-validation is shown in green.}
\label{fig:global_doppelganger}
\end{figure}

The number of principal components used in the compression, or encoding stage of RSCA (Step 1) is an important hyperparameter requiring tuning. In Figure \ref{fig:global_doppelganger}, we plot the doppelganger rate against the number of principal components, both with and without using the cross-validation procedure described in section \ref{sec:measuringchemicalsimilarity}. Results without cross-validation display significant overfitting and are mostly shown in an effort to highlight the importance of the cross-validation procedure.

This Figure illustrates how, unsurprisingly, RSCA's performance is strongly dependent on the PCA dimensionality.  Doppelganger rates decrease with increasing PCA dimensionality, up until a dimensionality of size 30. At K $>$ 30, doppelganger rates start increasing because of overfitting. Since on our studied dataset, RSCA reaches its peak performance for 30 PCA components, all further quoted results and figures will accordingly use the first 30 PCA components.

That RSCA's performance still improves up to a dimensionality of 30 is interesting. This demonstrates that a hyperplane of at least size 30 is required for capturing the intrinsic variations of APOGEE stellar spectra, a number noticeably larger than the 10 dimensional hyperplane found in \cite{pca_dimensionality}. Methodological differences between studies may partially explain such differences. For example, the PCA fit in \cite{pca_dimensionality} was applied on spectra displaying limited instrumental systematics and with non-chemical imprints on the spectra preliminary removed. It should also be noted that this does not mean that the chemical space is 30 dimensional. Some PCA dimensions may capture instrumental systematics or non-chemical factors of variation, such as residual sky absorption and emission and interstellar dust imprints. Also, since chemical species leave non-linear imprints, because of the linearity of PCA, each non-linear chemical dimension may require multiple PCA components to be fully captured.

\subsection{RSCA interpretability}

\begin{figure*}
\includegraphics[width=0.24\textwidth]{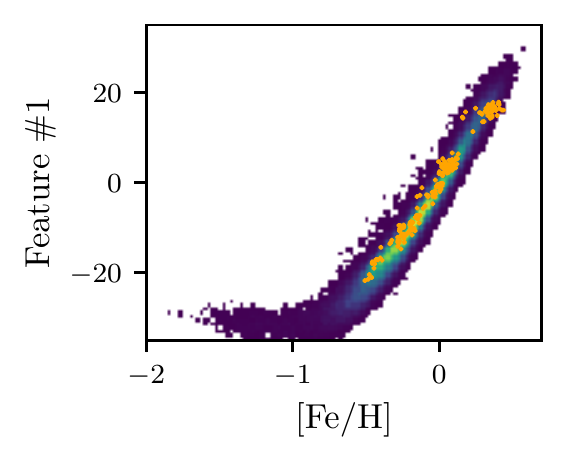}
\includegraphics[width=0.24\textwidth]{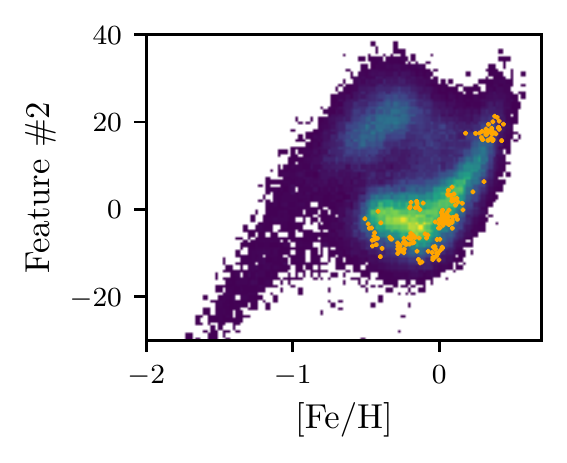}
\includegraphics[width=0.24\textwidth]{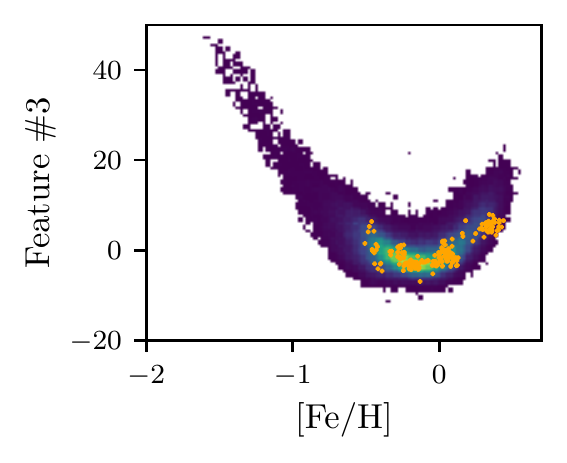}
\includegraphics[width=0.24\textwidth]{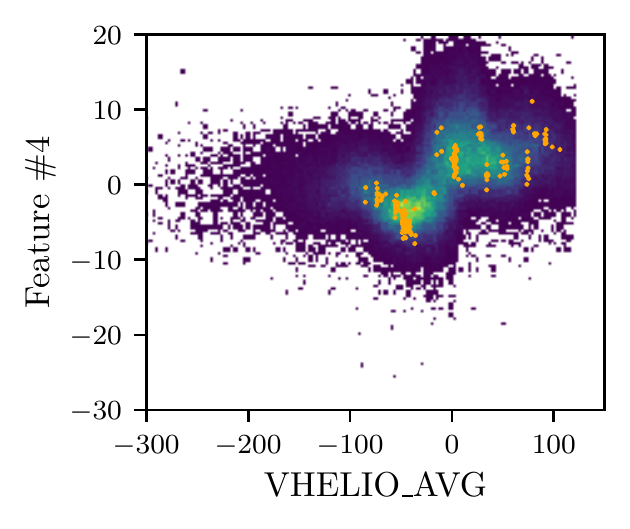}
\caption{Three features judged most important by metric learning approach plotted against [Fe/H] and [$\alpha$/Fe] for the 151,145 stars in $X_{pop}$ and the fourth most-important feature plotted against VHELIO\_AVG (radial velocity ASPCAP label). Location of the 185 stars in $X_{clust}$ (the open cluster dataset used to train the metric learning model) are shown by orange markers.}
\label{fig:interpretation}
\end{figure*}

We now study which spectral features are leveraged by the RSCA algorithm when recognizing open clusters. In the RSCA rescaled basis (Step 4), the dimensions are scaled proportionally to their perceived usefulness at measuring chemical similarity. Therefore, factors of variation judged most important by RSCA will correspond to the  most strongly scaled dimensions of the representation (ie with the largest $\frac{1}{\sigma_{r_i}}$). Figure \ref{fig:interpretation} shows the relationship between the three features with scaled dimensions with the largest amplitudes and metallicity, [Fe/H] and [$\alpha/Fe$] for a representation obtained by running the RSCA algorithm with a PCA dimensionality of 30.

As seen from the leftmost panel, there is a close relationship between "Feature \#1" - the RSCA dimension with the largest associated scaling factor and the ASPCAP [Fe/H] label. The relationship is close to a one-to-one mapping, which illustrates how this feature traces the metallicity content of the stellar spectra. Because "Feature \#1" (as a direction in a hyperplane of stellar spectra) is a linear function of the stellar spectra and metallicity is a non-linear feature, some degree of scatter in the relationship is expected.

The relationship between "Feature \#2" and [Fe/H] (second panel) exhibits the same bimodality as observed when plotting alpha enhancements [$\alpha$/Fe] against [Fe/H] \citep{Leung2018}. This indicates that "Feature \#2" captures $\alpha$-element enhancements. It is particularly noteworthy that we are able to recover the $\alpha$-element bimodality when the open clusters in our dataset (orange markers) are located only in the low-$\alpha$ sequence. This demonstrates the metric model's capacity to extrapolate to abundance values outside the range of values covered in the open cluster training dataset. This provides evidence that the model may still be effective for stars atypical of those in the open cluster dataset.

The relationship between "Feature \#3" and metallicity (third panel) is not as easily interpreted. Given the non-linear nature of metallicity, it is possible that it encodes residual metallicity variability not captured by "Feature \#1" but it is also possible that it contains some further independent chemical dimension.

This figure illustrates a nice property of RSCA. Because dimensions of the RSCA representation correspond to eigenvectors of the covariance matrix $\Sigma_{clust}$, the RSCA algorithm, at least to first order, assigns the distinct factors of variation within spectra to separate dimensions of the representation. That is, to say that the dimensions of the RSCA representation capture distinct factors of variation, such as the metallicity or the alpha-element abundance, rather than a combination of factors of variation. Additionally, the most important factors of variation for recognzing open clusters occupy the dimensions with the largest scaling factors. This property makes RSCA particularly versatile. For example RSCA can be used to separate out high and low  $\alpha$-abundance stars in the disk, or to select low-metallicity stars. Additionally, it is likely that because of this property RSCA could be used to search for hidden chemical factors of variation within stellar spectra, although this has not been attempted in this paper.

We found that some of the dimensions of the RSCA representation showed trends with radial velocity. An example of a dimension showing a trend with radial velocity is shown in the last panel of Figure \ref{fig:interpretation} and an investigation into the detailed causes of the radial velocity trends is presented in Appendix \ref{appendix:radial}. The existence of such trends indicates that even after our pseudo-continuum normalization procedure, RSCA is still capable, at least weakly, to exploit radial velocity information in the spectra to recognize stellar siblings. Because only a subset of the dimensions show such trends, a representation tracing only chemistry can be obtained by only keeping those dimensions which show no trends with radial velocity. In this work, we propose to only keep the first three dimensions of the representation. While this choice may appear particularly stringent, as we will show in coming sections, these three dimensions contain the bulk of the discriminative power of the representation (see Table \ref{tab:dimensionality}).

\subsection{Comparison of using RSCA versus measured abundances in calculating chemical likeness}\label{sec:results}

\begin{figure}
\includegraphics[width=\columnwidth]{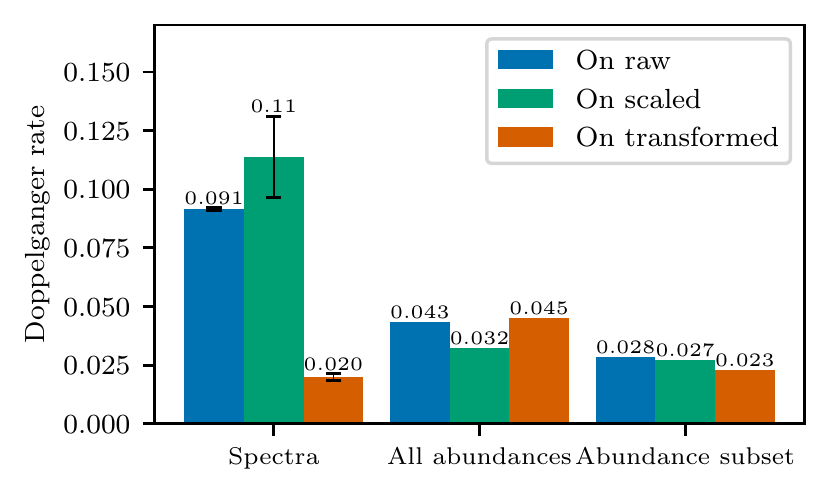}
\caption{Global doppelganger rates estimated for varying metric-learning approaches and representations. On the x-axis, ``spectra'' refers to doppelganger rates obtained from spectra $X$ after dimensionality reduction with PPCA to a 30-dimensional space, ``all abundances" to doppelganger rates obtained from a representation formed from the full set of APOGEE abundances in $Y$, ``abundance subset" to doppelganger rates obtained using a representation formed only from the abundances for the following species: Fe,Mg,Ni,Si,Al,C,N. Global doppelganger rates "on raw" (blue) are obtained by measuring distances in the raw representation without any transformation of the representation, ``on scaled'' (green) are obtained by applying the scaling transform on the raw representation without preliminary application of the sphering and reparametrization transform (Steps 1 and 4 for spectra and only Step 4 for abundances which do not need dimensionality reduction), "on transformed" are obtained by applying all steps of the proposed metric learning approach (Steps 1,2,3 and 4 for spectra and Steps 2,3,4 for abundances). As the implementation of the PPCA algorithm used in this paper yielded stochastic PCA components, doppelganger rates from spectra correspond to the mean across 10 runs with error bars corresponding to the standard deviation amongst runs.} 
\label{fig:ablation}
\end{figure}

In this section, we compare the effectiveness of measuring chemical similarity using data-driven approach, with that achievable from using measured stellar abundances. To do so, we compare the doppelganger rates that are obtained by RSCA to those from using stellar abundance labels. The results of such a comparison are shown in Figure \ref{fig:ablation}. For this Figure, doppelganger rates are measured from abundances consistently with the RSCA approach (see Figure caption for more detail), such that any differences in performance can be attributed to underlying differences in the information content of the representation. We remind the reader that the doppelganger rates are evaluated for pairs of stars at the same extinction and radial velocity.  This guarantees that the doppelganger rate cannot be artificially reduced through our model exploiting information relating to radial velocity or extinction. Per-cluster doppelganger rates are also provided in Appendix \ref{appendix:local}.

From this Figure, we see that executing all steps of the RSCA is crucial for obtaining low doppelganger rates when working directly with spectra. This is seen from how the low doppelganger rates are only obtained with full application of our metric-learning approach (red). On the other hand, when working with stellar abundances, the RSCA approach appears to bring only limited benefits - as seen from the only slight difference in doppelganger rates between measuring distances in the raw abundance space (blue) and in the transformed space (red) that is obtained by application of our full-metric learning approach minus the PCA compression. This result is not surprising and reflects how most steps of the RSCA approach are designed with the aim of generating a representation comparable to stellar labels. That is, to say one where all factors of variation other than chemical factors of variation are removed. The RSCA approach does however still bring some benefits when working with abundances as seen from the lower doppelganger rates.

This Figure also shows that excluding chemical species improves the doppelganger rate. This is seen from how the doppelganger rate is lower when using a carefully chosen subset of species (right) than when using the full set of abundances (center). This can appear counter-intuitive as it implies that more data leads to worsened performance but in this specific case, where the uncertainties on abundances are not accounted for, it can be justified by the low intrinsic dimensionality of chemical space. Since many species contain essentially the same information, adding species with higher uncertainty into the representation adds noise into the representation. It does this without contributing any additional information beneficial for recognizing open clusters. We expect that such an effect would disappear when accounting for uncertainties on stellar labels, but it is still a good illustration of the brittleness of abundance-based chemical tagging.

The combination of species shown in red is the set of species which were found, after manual investigation, to yield the lowest doppelganger rates. This is the combination of stellar individual element abundance labels Fe,Mg,Ni,Si,Al,C,N (with respect to Fe with the exception of Fe which is with respect to H). The doppelganger rate from this combination of species, of 0.023, despite being the smallest doppelganger rate achieved from stellar labels, is higher than the doppelganger rate obtained from stellar spectra of 0.020 (2 percent). That our method is able to produce better doppelganger rates from spectra than from stellar labels highlights the existence of information within stellar spectra not being adequately captured by stellar labels. While stellar labels are derived from synthetic spectra which only approximately replicate observations, our fully data-driven model makes direct use of the spectra translating into lower doppelganger rates.

\begin{figure}
\includegraphics[]{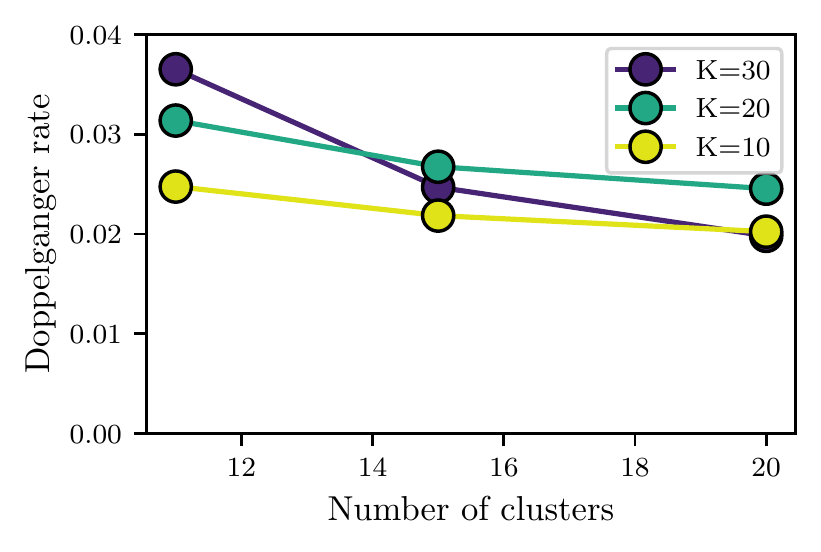}
\caption{Expected global doppelganger rates when training a metric-learning model on only a subset of all open clusters in $X_{clust}$ with a number of clusters given by the x-axis. Results for different PCA dimensionalities used for compressing stellar spectra are represented by different colored lines. Clusters used in the expectated doppelganger rate calculations were chosen randomly from $X_{clust}$, and quoted results are for the average of 50 repeated trials.}
\label{fig:datasize}
\end{figure}

\subsection{Dimensionality of chemical space}

That the PCA representation is 30 dimensional does not mean that all 30 dimensions carry information useful for recognizing open clusters. To get a grasp of the dimensionality of the chemical space captured by RSCA, we calculated the doppelganger rates for RSCA representations in which only the dimensions with the largest scaling factors are kept (ie other dimensions are excluded from the doppelganger rate distance calculations). We calculate doppelganger rates multiple times, each with a different number of dimensions preserved. The results of this investigation are shown in Table \ref{tab:dimensionality}. From this Table, we see that the dimensionality of spectra appears to be, at least to first degree, extremely low. The top two dimensions of the RSCA model (as shown in Figure \ref{fig:interpretation}) are capable of matching the performance obtained from stellar labels whilst the top four dimensions exceed the performance from using the full representation. The four-dimensional representation is even more effective at recognizing chemically identical stars than the full RSCA representation which itself was more effective than stellar labels. It is not a new result that the dimensionality of chemical space probed by APOGEE is low. Recent research suggest that, at the precisions captured by APOGEE labels, chemical abundances live in a very low-dimensional space, for stars of the disk. For example, it was found in \cite{Ness2019,HowManyElements} that [Fe/H] and stellar age could predict all other elemental abundances to within or close to measurement precision. However, while previous analysis have depended on abundances to show this, here we can do this directly from spectra. As our methodology directly picks up on factors of variation and, if not controlled for, is capable of picking up on weak factors of variation such as diffuse interstellar bands, we can be confident that any remaining chemical factors of variation are either i) highly non-linear for our model to not be capable of picking up on them, ii) very weak spectral features, or iii) not particularly discriminative of open clusters as would be the case for chemical variations arising from internal stellar processes for example induced by accretion of planetary materials or internal stellar processes.

\begin{table}
\centering
\begin{tabular}{p{1.5cm} p{3.0cm}}
 \hline
N & Doppelganger rate \\
 \hline
1 & $0.0962 \pm 0.0212$ \\
2 & $0.0219 \pm 0.0025$ \\
3 & $0.0198 \pm 0.0028$ \\
4 & $0.0182 \pm 0.0021$ \\
5 & $0.0180 \pm 0.0021$ \\
6 & $0.0188 \pm 0.0017$ \\
7 & $0.0184 \pm 0.0017$ \\
30 & $0.0199 \pm 0.0015$ \\
\end{tabular}
        \caption{Global doppelganger rate obtained by the RSCA model applied to stellar spectra in which all but the N most strongly scaled dimensions of a 30 dimensional RSCA representation are discarded. As the implementation of the PPCA algorithm used in this paper yielded stochastic PCA components, doppelganger rates from spectra correspond to the mean across 10 runs with error bars corresponding to the standard deviation amongst runs.}
        \label{tab:dimensionality}
\end{table}

\subsection{Impact of Dataset Size}

Our method learns to measure chemical similarity directly from open clusters without reliance on external information. Because of this, its performance will be tightly linked to the quality and quantity of data available. Figure \ref{fig:datasize} attempts to estimate our method's dependency on the size of the open cluster dataset. In this figure, for varying PCA dimensionalities, we plot the expected doppelganger rate for an open cluster dataset containing a given quantity of open clusters, whose number is given by the x-axis. We estimate the expected doppelganger rate for a given number of open clusters by estimating and averaging the doppelganger rates for all data subsets containing that number of clusters. From this figure, we see that the larger PCA dimensionalities still benefit from the addition of open clusters. This is suggestive that performance would likely further improve with access to additional open clusters. Such larger datasets may also enable the usage of more complex non-linear metric-learning approaches which may yield further improvements not captured in this figure.

\section{Discussion}\label{sec:discussion}

We have presented a novel approach for identifying chemically similar stars from spectroscopy based on training a metric-learning model on the spectra of open cluster stars. This approach has several appealing properties. It is end-to-end data-driven in the sense that it does not rely on synthetic models nor labels derived from synthetic models. This method only makes use of open cluster members, which can themselves be identified with minimal reliance on theoretically-derived quantities \citep{openclusteridentification1,openclusteridentification2,openclusteridentification3}. This makes the method insensitive to the domain gap of synthetic spectra. Additionally, where traditional spectral fitting approaches require instrumental systematics to be fully suppressed, lest they further exacerbate the domain gap, our fully data-driven approach, at least in theory, automatically learns on its own to ignore most variations due to instrumental systematics.

 We expect that the approach that we have developed will perform particularly well in a number of regimes. For example,  low resolution spectra, where blended features lead to compounding model inaccuracies. Additionally, for  M-type stars, where molecular features complicate the retrieval process. In general, our method will likely be an efficient and effective approach where theoretical models are inaccurate or the observed spectra itself is plagued by complex systematics. This is for example the case for M dwarfs \citep{Birky_2020,Behmard_2019}.

Although our data-driven algorithm shows excellent performance, one may wonder whether there is still room for further improvements, particularly since strong chemical tagging, if ever possible, will require improvements in our chemical similarity measurements \citep{HowManyElements}. There are reasons to be hopeful here. 

Firstly, our data-driven model comes with clear limitations which are not inherent to the approach, but rather imposed by our modelling choices. There are many algorithmic choices for building a metric learning model optimized at distinguishing open clusters \citep[e.g.][]{LargeMarginNearestNeighborClassification,Neighbourhood,relevantComponentAnalysis,pml1Book} and ours is only one of many choices. In particular, the ability of RSCA to extract the chemical content of stellar spectra is constrained by its linear nature. Although this linearity is convenient for avoiding over-fitting, enabling better out-of-distribution performance and cross-validation, it also artificially limits the precision with which ``pure" features containing only chemical signal can be learned. One can be hopeful that a suitably regularized non-linear metric learning models, such as for example twin neural network \citep{siamese,pml1Book}, could surpass our model. However, building such a model from our limited number of open clusters would present its own unique challenges.

Secondly, by being entirely data-driven, the performance of this approach is inexorably linked to the quality and quantity of data available. This makes it poised to benefit from new open cluster discoveries and/or deliberately targeted observations. Improvements may also be possible by leveraging other source of chemically similar stars such as wide binaries.

Our method also comes with caveats. Data-driven methods do not extrapolate well outside of their training dataset. As such, performance may be lower for clusters that are atypical compared to those in the open cluster reference set. Since open clusters are typically younger stars \citep{openclusterreview}, this means performance may be decreased on older cluster stars. However, given the tight relationships between RSCA dimensions and chemical parameters for stars in $X_{pop}$, such an effect is likely to be small. Additionally, our model makes no use of the error information in spectra, which is valuable information that could likely be squeezed-out for even better performance.

Another downside of our approach is its coarse-grained nature. While stellar labels provide a fine-grained view into chemical similarity with a breakdown into chemical composition of individual species, our approach only provides a coarse-grained  measurement of chemical similarity. This limits the types of scientific problems the approach can be used to answer. However, it may very well be possible to extend the method from measuring overall chemical similarity to measuring individual elemental abundances. A way in which this could possibly be done is through applying it to windows centered on the locations of stellar lines instead of to the full spectrum. Also, it is not always clear what exact information is captured by the representation, and in particular there is always a risk, despite all of our checks, that the model is acting on non-chemical information within the spectra.

\section{Conclusion}
Large-scale Galactic surveys, the likes of APOGEE, LAMOST and GALAH, have collected hundreds of thousands of high-quality stellar spectra across the galaxy. These surveys are vastly broadening our understanding of the Milky Way.  How best to analyse these spectra however still remains an open-question. One limitation is that traditional spectral fitting methods currently do not make full use of the information in stellar spectra. This is largely because our stellar models are approximations.

In this paper we developed a fully data-driven, linear metric learning algorithm operating on spectra for extracting the chemical information within stellar families. Through experiments on APOGEE, we demonstrated that our metric learning model identifies stars within open clusters more precisely compared to using stellar labels which indicates an improved ability to discriminate between chemically identical stars. We further found that our model's capacity to distinguish open clusters could largely be attributed to a two-dimensional subspace of our final representation which was found to approximately coincide with metallicity and $\alpha$-elemental abundances. That our model's capacity at recognizing open clusters plateaus at N$\sim$ 4 supports the idea that the dimensionality of chemical space probed by APOGEE is, for Galactic archaeology purposes, low, in the disk. However, we do find hints of further dimensions potentially containing chemical information. 

There are several reasons why our metric-learning approach could be favoured over using stellar labels. It can be applied to spectra of stars for which we do not yet have very good synthetic spectra and so would otherwise not be able to analyze well. It is completely independent of our theoretical knowledge of stellar atmospheres and so could be used to validate existing astronomical results in a way which is independent of any biases which may exist in our synthetic spectra. Finally and perhaps most importantly, whereas the traditional derivation of stellar labels is fundamentally limited by our inability to generate faithful synthetic spectra, our metric learning approach does not suffer from such a limitation. This means that by improving the quality of the training dataset and the metric-learning approach used, performance may be further improved.

\begin{acknowledgements}
DDM is supported by the STFC UCL Centre for Doctoral Training in Data Intensive Science (grant number ST/P006736/1). DDM thanks Serena Viti for helpful discussions.

Melissa K Ness is supported in part by a Sloan Foundation Fellowship.

Funding for the Sloan Digital Sky 
Survey IV has been provided by the 
Alfred P. Sloan Foundation, the U.S. 
Department of Energy Office of 
Science, and the Participating 
Institutions. 

SDSS-IV acknowledges support and 
resources from the Center for High 
Performance Computing  at the 
University of Utah. The SDSS 
website is www.sdss.org.

SDSS-IV is managed by the 
Astrophysical Research Consortium 
for the Participating Institutions 
of the SDSS Collaboration including 
the Brazilian Participation Group, 
the Carnegie Institution for Science, 
Carnegie Mellon University, Center for 
Astrophysics | Harvard \& 
Smithsonian, the Chilean Participation 
Group, the French Participation Group, 
Instituto de Astrof\'isica de 
Canarias, The Johns Hopkins 
University, Kavli Institute for the 
Physics and Mathematics of the 
Universe (IPMU) / University of 
Tokyo, the Korean Participation Group, 
Lawrence Berkeley National Laboratory, 
Leibniz Institut f\"ur Astrophysik 
Potsdam (AIP),  Max-Planck-Institut 
f\"ur Astronomie (MPIA Heidelberg), 
Max-Planck-Institut f\"ur 
Astrophysik (MPA Garching), 
Max-Planck-Institut f\"ur 
Extraterrestrische Physik (MPE), 
National Astronomical Observatories of 
China, New Mexico State University, 
New York University, University of 
Notre Dame, Observat\'ario 
Nacional / MCTI, The Ohio State 
University, Pennsylvania State 
University, Shanghai 
Astronomical Observatory, United 
Kingdom Participation Group, 
Universidad Nacional Aut\'onoma 
de M\'exico, University of Arizona, 
University of Colorado Boulder, 
University of Oxford, University of 
Portsmouth, University of Utah, 
University of Virginia, University 
of Washington, University of 
Wisconsin, Vanderbilt University, 
and Yale University.

\end{acknowledgements}

\appendix
\section{Interstellar masking} \label{appendix:masking}
Regions containing interstellar absorption features are identified from the APOGEE data using a data driven-procedure as described below.

The method makes use of two datasets: one containing spectra of stars at low extinction $X_{low}$ which should not contain interstellar features and one of high extinction stars $X_{high}$ which should contain strong interstellar features. $X_{low}$ is formed through a dataset cut on $X_{pop}$ in which only stars with AK\_TARG$<$ 0.005 are kept, $X_{high}$ only preserves stars with AK\_TARG$>$ 0.5. 

We apply PCA with 30 principal components to the dataset of low extinction stars $X_{low}$ to obtain a PCA basis capturing the natural variations amongst stellar spectra at low extinction.  Since this PCA basis only captures the variations amongst low extinction stars, features in high-extinction stars associated to the interstellar medium will be poorly reconstructed by the projection onto this low-extinction PCA hyperplane.

\begin{figure*}
\includegraphics[width=\linewidth]{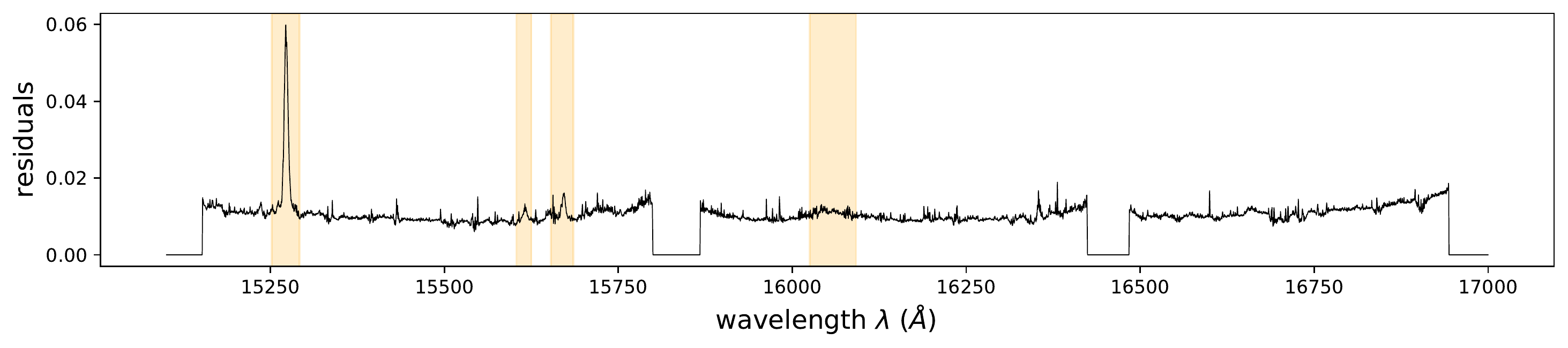}
\caption{Mean residual per wavelength between stellar spectrum and their projection on the low-extinction PCA hyperplane averaged over all stars in the high-extinction dataset $X_{high}$. High residual spectral bins correspond to wavelengths where interstellar extinction strongly affects stellar spectra. Region highlighted in yellow are the regions that were chosen to be censored to suppress interstellar features from the spectra.}
\label{fig:diffuse_interstellar}
\end{figure*}

In Figure \ref{fig:diffuse_interstellar}, we plot the mean residual per wavelength between stellar spectrum and their projection on the low-extinction PCA hyperplane averaged over all stars in the high-extinction dataset $X_{high}$. High-residual regions in this fit will correspond to regions poorly captured by the low-extinction PCA hyperplane. Comparing the high-residual regions with the locations of known diffuse interstellar bands reveals excellent agreement \citep{diffuse1,diffuse2}. For this paper, those regions that should be censored were selected manually with the final choice of regions overlain in yellow in Figure \ref{fig:diffuse_interstellar}.

\section{Visualizing radial velocity instrumental systematics} \label{appendix:radial}

\begin{figure*}
\includegraphics[width=\linewidth]{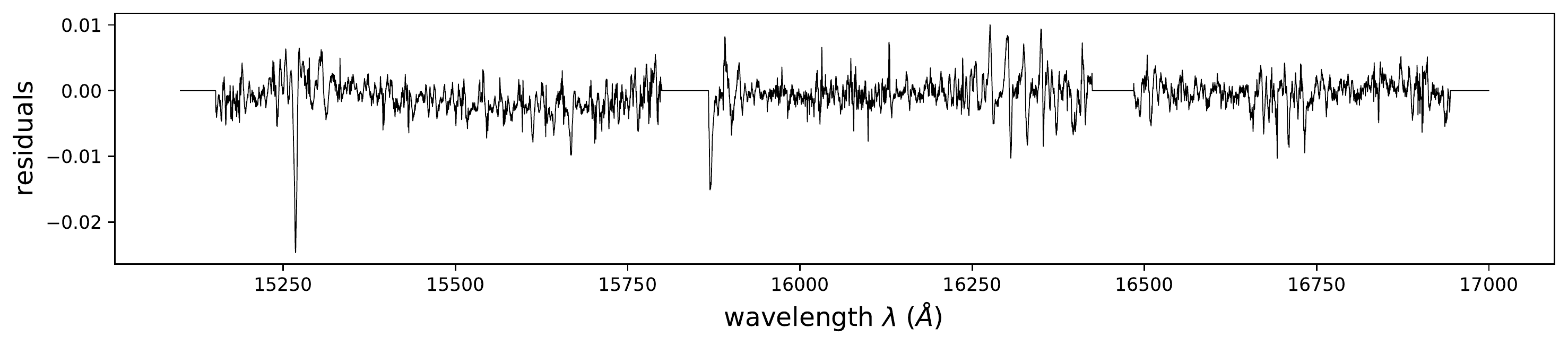}
\caption{Mean residual per wavelength between stellar spectrum and their projection on the low radial-velocity PCA hyperplane averaged over all stars in the high radial-velocity dataset $X_{high}$. High residual spectral bins correspond to spectral regions with strong dependence on radial velocity.}
\label{fig:radial_velo}
\end{figure*}

We apply a similar approach to that taken for extinction in Appendix \ref{appendix:masking} to radial velocities. That is to say we create a dataset of low radial-velocity stars by selecting only stars for which $|\mbox{VHELIO\_AVG}|<5$~kms$^{-1}$. We train a PCA model on the low-velocity spectra and visualize the PCA model's residuals on a dataset of high-velocity spectra with $\rm VHELIO\_AVG>80$~kms$^{-1}$. As in the previous section we use a 30 dimensional PCA model.

In Figure \ref{fig:radial_velo}, we plot the mean absolute residual per wavelength between stellar spectrum and their projection on the low-velocity PCA hyperplane averaged over all stars in the high radial-velocity dataset $X_{high}$. High-residual regions in this fit will correspond to regions poorly captured by the low-velocity PCA hyperplane. In this plot, in addition to the diffuse interstellar bands, there appears to be other regions with weak instrumental systematics correlated with radial velocity.

It is worth mentioning that a variant of RSCA can be used for removing radial velocity imprints on the spectra. By applying the RSCA algorithm to same radial velocity stellar groups instead of open clusters, one can identify a hyperplane of the stellar spectral space capturing solely spectral features correlated to radial-velocity. Substracting variations within this hyperplane from stellar spectra then yields spectra in which features correlated with radial-velocity are selectively suppressed. Here, we do not apply such a preprocessing procedure as it complicates the analysis while not improving over the simpler procedure of only keeping the first three dimensions.

\section{Checking for instrumental systematics}\label{appendix:leakage}

It is worthwhile to ascertain that our model, when identifying open clusters, is only relying on chemical features within the spectra and not on instrumental systematics that happen to be predictive of open clusters and would not transfer towards identifying dissolved clusters. This is especially important as any dependency on instrumental systematics would lead to overly optimistic doppelganger rates.

Because the stars in open clusters are gravitationally bound, they are often part of the same telescope field of view and so observed simultaneously on nearby fibers of a plate. Being observed together could plausibly introduce systematics (due to instrumental imperfections or telluric residuals) which could then be actioned on by the metric learning-model when identifying open clusters. Here, we run an experiment to ascertain that this is not an issue for our model.

Since any such shared instrumental systematic will only affect stars observed simultaneously, we can validate that our model is not exploiting shared instrumental systematics by comparing the similarity distributions for stellar siblings that were observed together on the same plate and for stellar siblings that were not. The idea being that a model exploiting instrumental systematics would have a lower doppelganger rate on the pairs of stars observed together than on the pairs of stars observed separately.

To separate stellar siblings into pairs of stars observed together and pairs observed separately, we went through all pairs of stellar siblings in the open cluster dataset. Using the individual observation dates of exposures that comprise the combined spectra, as provided by the VISITS allStar field, we categorized pairs of siblings into two groups: those pairs composed of stars observed together i.e. with the same dates of visits for exposures, and those pairs of stars observed separately. When doing this analysis we discarded the small fraction of stellar pairs for which not visit dates only partially overlapped.

\begin{figure*}[htb!]
\includegraphics[width=\textwidth]{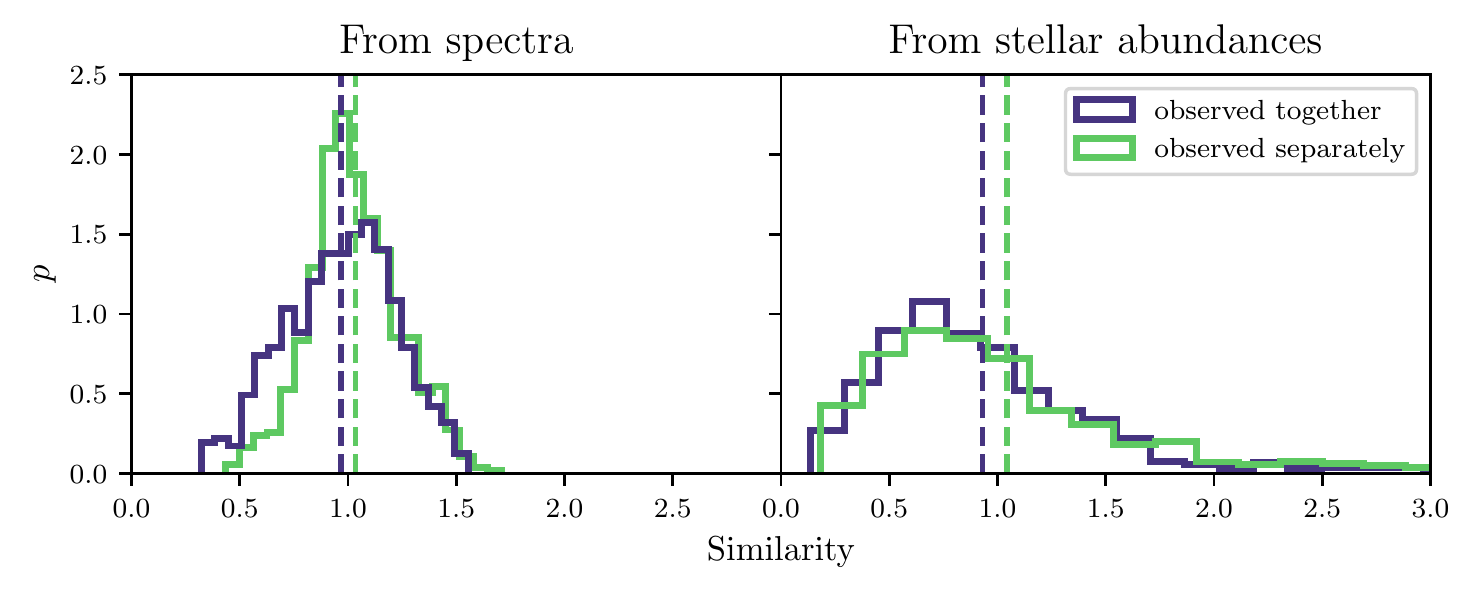}

\caption{Investigation into metric-learning models dependency on instrumental systematics. "From masked spectra" refers to distances derived from a metric-learning model applied to masked stellar spectra. "Stellar abundances" refers to distances derived from a core set of abundances (see Section \ref{sec:experiments} for full details).}
\label{fig:validation_bis}

\end{figure*}

In Figure \ref{fig:validation_bis}, we show the distributions of chemical similarities for pairs of open cluster stars observed simultaneously compared to pairs observed separately. At left, we show our metric learning approach applied to spectra (with a 30 dimensional latent). At right, we show this applied to abundances. For both metric-learning models, stars observed together are predicted to be slightly more chemically similar than stars observed separately. Since, both approaches, applied to spectra and to abundances,  provide similar behaviours with visit overlap, we conclude that our method is not making strong use of instrumental systematics when recognizing open clusters. It is nonetheless interesting that both approaches seem to marginally favour stars observed together as being more chemically similar, although given the small number of open clusters may be due to small sample sizes.

\section{RSCA Pseudocode} \label{appendix:pseudocode}

The Pseudocode for the RSCA algorithm.

\begin{figure*}[htb!]
    \centering
\begin{algorithm*}[H]
\SetAlgoLined
\KwData{
 $X_{\rm clust}$,
 $X_{\rm pop}$\,
 }
 $\rm compressor = \rm PPCA(\rm data =X_{pop},n_{\rm components} = N_K)$ \tcp*{Step 1}
 $Z_{\rm pop} = \rm compressor.transform(X_{pop})$\;
 $Z_{\rm clust} = \rm compressor.transform(X_{clust})$\;
 $\rm spherer = \rm sphere(\rm data =Z_{pop})$ \tcp*{Step 2}

 $Z_{\rm pop} = \rm spherer.transform(Z_{\rm pop})$\;
 $Z_{\rm clust} = \rm spherer.transform(Z_{clust})$\;
 $Z_{\rm intra-cluster} = \rm ZeroCenterClusters(Z_{clust})$ \tcp*{Step 3}
 $\rm reparametrizer = \rm PCA(\rm data =Z_{\rm intra-cluster} ,n_{\rm components} = N_K)$ \;
 $Z_{\rm pop} = \rm reparametrizer.transform(Z_{pop})$\;
 $Z_{\rm clust} = \rm reparametrizer.transform(Z_{clust})$\;
 \For{ \rm i = 1 to $N_K$}
 {
 $\sigma_{\rm clust_i}^{2}=\frac{\sum_{j=1}^{k}\left(n_{j}-1\right) \sigma_{ji}^{2}}{\sum_{j=1}^{k}\left(n_{j}-1\right)}$ \tcp*{Step 4}
 $\sigma_{\rm pop_i}^{2}=1$\ \tcp*{because of sphering}
 $\sigma_{\rm r_{i}}= \frac{\sigma_{\rm clust_{i}}\sigma_{\rm pop_{i}}}{\sqrt{\sigma_{\rm pop_{i}}^2-\sigma_{\rm clust_{i}}^2}}$\; 
 $Z_{\rm pop_i} = Z_{\rm pop_i}\div \sigma_{\rm r_{i}}$\;
 $Z_{\rm clust_i} = Z_{\rm clust_i}\div \sigma_{\rm r_{i}}$
 }
\caption{RSCA Algorithm}
\end{algorithm*}
\label{fig:pseudocode}
\end{figure*}

\section{Per-cluster Doppelganger Rates}\label{appendix:local}

The doppelganger rates for all open clusters in $X_{clust}$ (Figures \ref{fig:dop_local1} \& \ref{fig:dop_local2})

\begin{figure}
\includegraphics[width=0.48\textwidth]{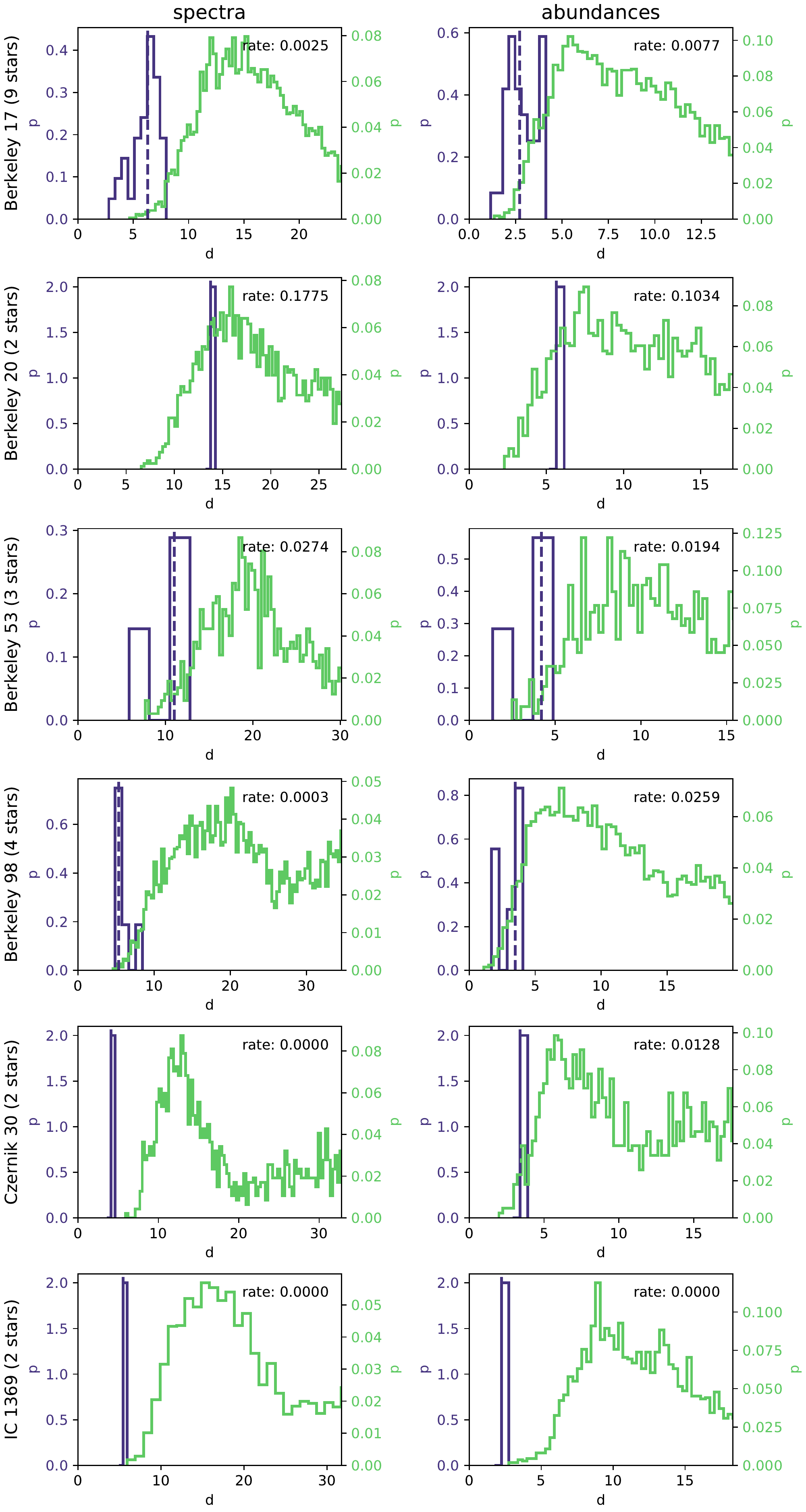}
\hspace{0.04\textwidth}
\includegraphics[width=0.48\textwidth]{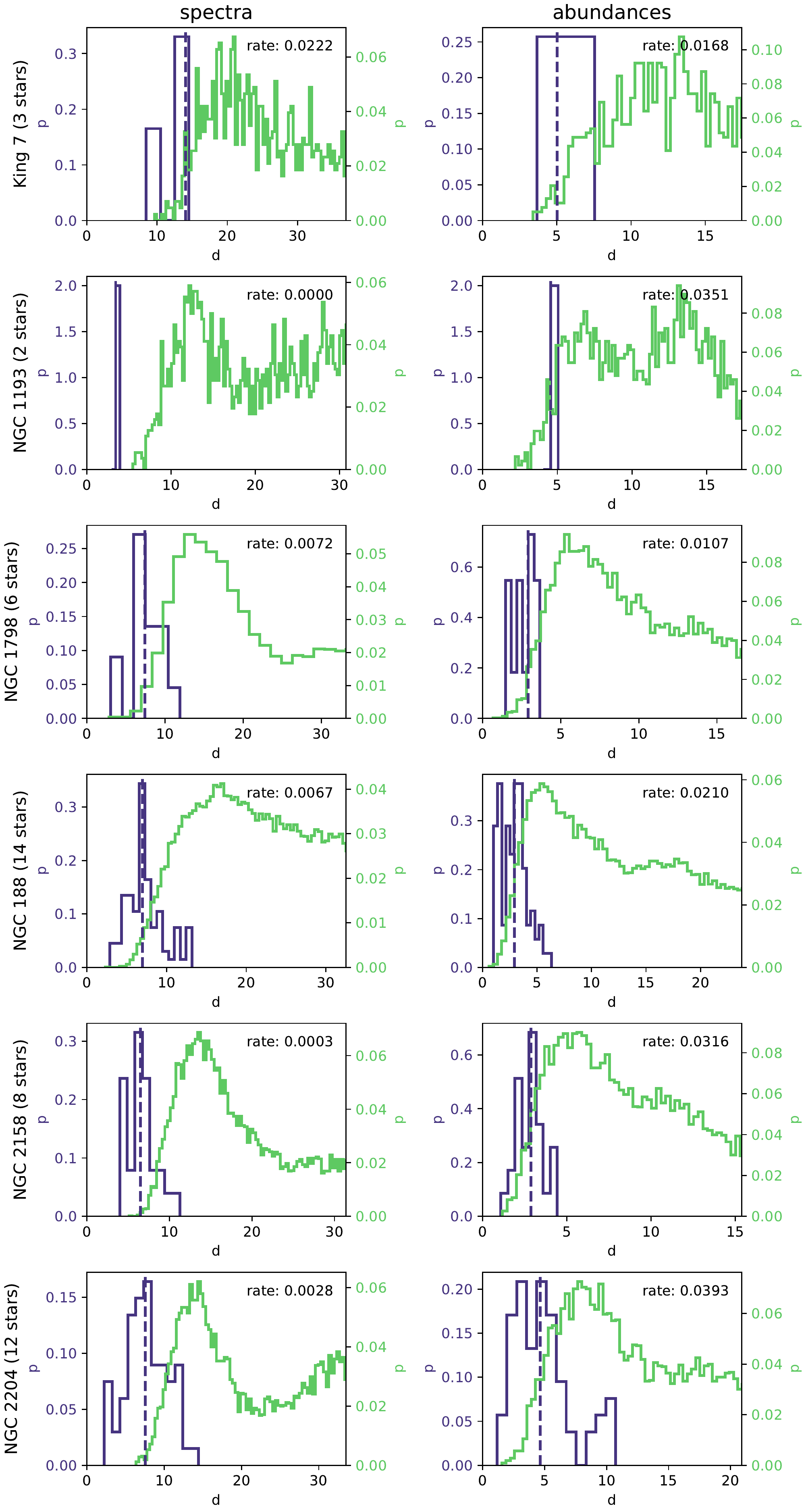}
\caption{Histograms of the chemical similarity between open cluster pairs of stars as predicted by the metric-learning approach. For each open cluster, the distribution of inter-cluster similarities, calculated as the distribution of similarities between pairs of stars composed of one random cluster member and a random field star, is shown in green and the distribution of intra-cluster similarities - similarity between pairs of stellar siblings - is shown in blue. The median intra-cluster similarity, as used in doppelganger rate calculations, is marked by dashed vertical line. The leftmost panel displays the histograms derived from applying the metric-learning approach to stellar spectra. The rightmost panel displays the histograms derived from applying the metric-learning approach to the "abundance subset" as defined and described in Section \ref{sec:results}. Doppelganger rates for individual clusters are shown in top-left corner of every panel.}
\label{fig:dop_local1}

\end{figure}
\clearpage

\begin{figure}
\includegraphics[width=0.45\textwidth]{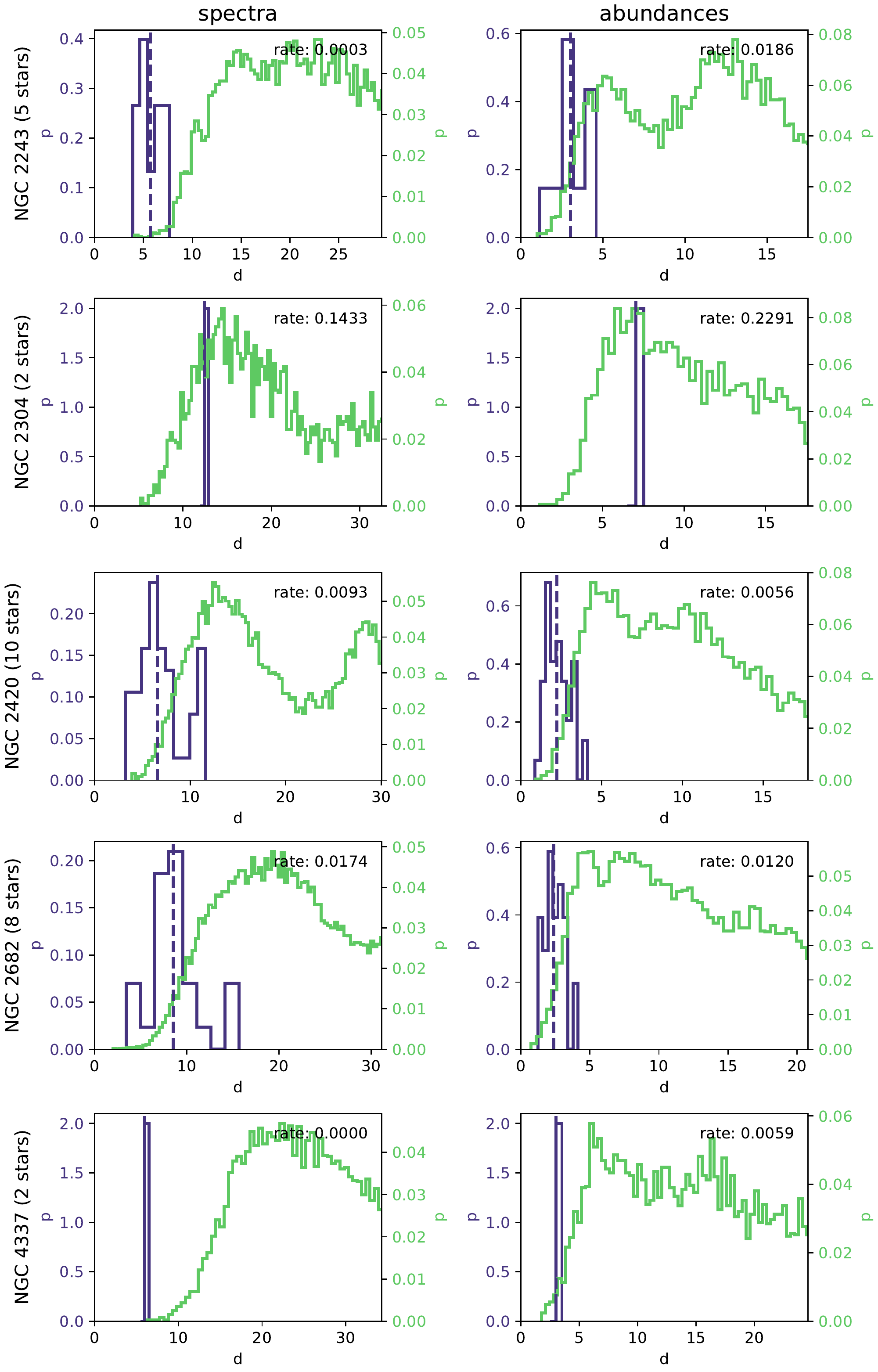}
\hspace{0.1\textwidth}
\includegraphics[width=0.45\textwidth]{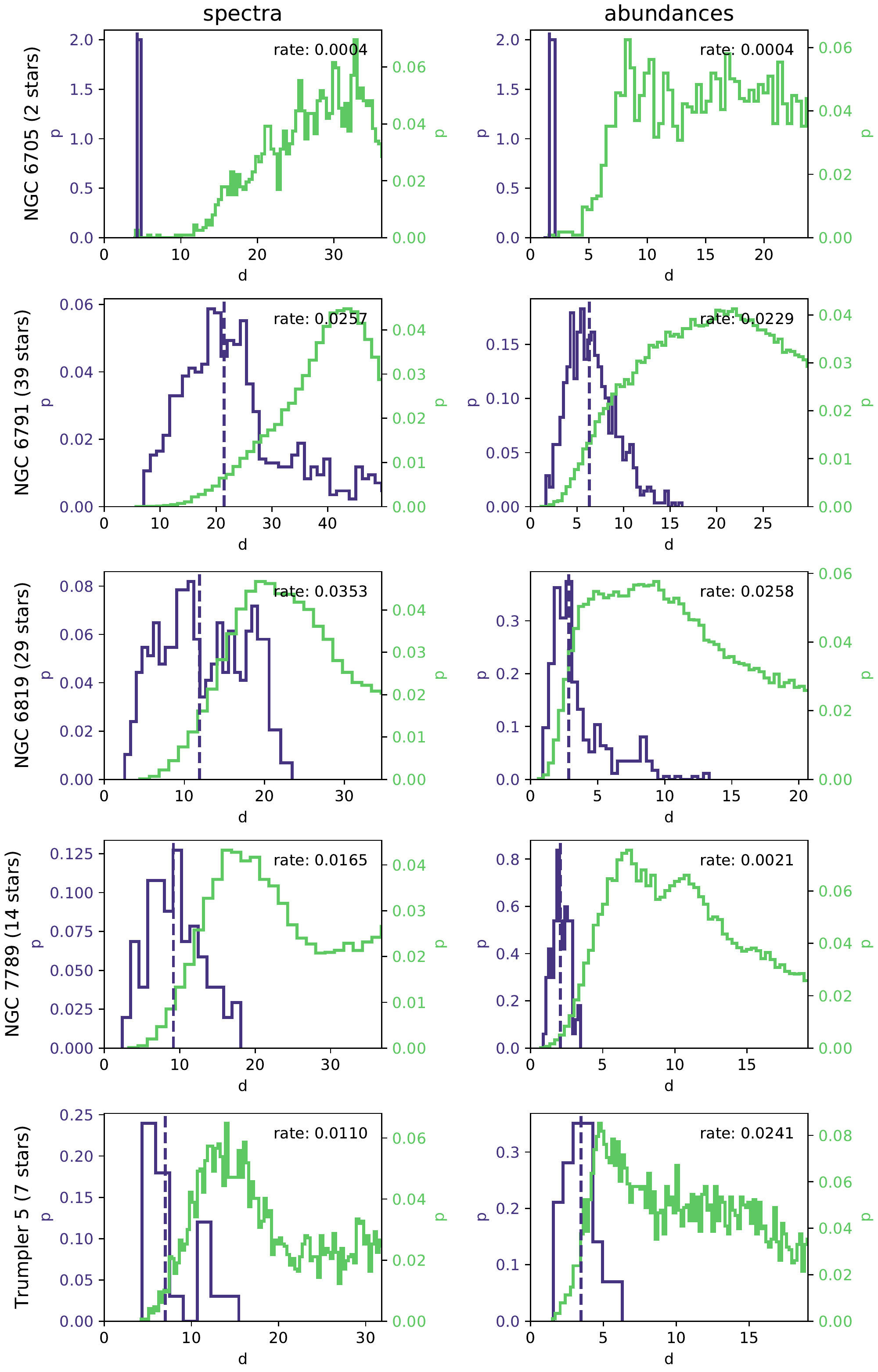}
\caption{Continuation of Figure  \ref{fig:dop_local1}.}
\label{fig:dop_local2}
\end{figure}
\clearpage


\bibliography{references}{}
\bibliographystyle{aasjournal}




\end{document}